\documentclass[a4paper,11point]{scrartcl}
\usepackage{hyperref}
\usepackage{epsfig}
\usepackage{times}
\usepackage{amsmath}
\usepackage{amsbsy}
\usepackage{amsfonts}
\usepackage{multicol}
\usepackage[latin1]{inputenc}
\usepackage{url}
\typearea{12}
\begin{document}                % INITIALIZE - DONT CHANGE
\title{ A General Method for the Determination of Matrix Coefficients for High Order Optical System Modelling }
\author{\textbf{J. B. Almeida}\\ \normalsize
{Universidade do Minho, Departamento de F\'isica,}\\\normalsize
{4710-057 Braga, Portugal.}}  %
\date{July 1998}
\maketitle
%\begin{center}
%\today
%\end{center}
\begin{abstract}                % DON'T CHANGE THIS LINE
The non-linear transformations incurred by the rays in an optical system can be suitably described by matrices to any desired order of approximation. In systems composed of uniform refractive index elements, each individual ray refraction or translation has an associated matrix and a succession of transformations correspond to the product of the respective matrices. This paper describes a general method to find the matrix coefficients for translation and surface refraction irrespective of the surface shape or the order of approximation. The choice of coordinates is unusual as the orientation of the ray is characterised by the direction cosines, rather than slopes; this is shown to greatly simplify and generalise coefficient calculation. Two examples are shown in order to demonstrate the power of the method: The first is the determination of seventh order coefficients for spherical surfaces and the second is the determination of third order coefficients for a toroidal surface.
\end{abstract}

\section*{Keywords} Optics, Aberration, Matrices.
\section{Introduction}

An optical system can be effectively modelled in paraxial approximation by a product of  $4 \times 4$ matrices, each representing one elementary transformation of the light rays \cite{Gerrard}; the elementary transformations are either translations of the ray in homogeneous media or the effects of surfaces separating different media. A more accurate approach implies the consideration of higher order terms but the fact that Snell's law makes use of the sine function rules out the terms of even order; as a consequence, when one wants to improve on the paraxial approximation, one has to consider third order terms.

Aberrations have already been studied extensively \cite{Born,Slyusarev} but work is still going on in order to design symbolic models of optical systems that computers can use for optimisation purposes and humans can look at to gain a better understanding of systems' performance.

The matrix theory has been extended to deal with higher order terms \cite{Kondo} through the use of a vector basis that incorporates two position and two orientation coordinates as well as all their third or higher order monomials, increasing the overall dimension which becomes 24 for third order approximation. It is possible to apply axis symmetry to reduce the matrix dimension through the use of complex coordinates and their higher order monomials; for instance, in third order the matrices to be considered for axis-symmetric systems are $8 \times 8$ \cite{Kondo,Laks}.

The set of four coordinates normally used to describe the ray consists of the two rectangular coordinates, $x$ and $y$ along with the two ray slopes $u={\rm d}x /{\rm d}z$ and $v={\rm d} y / {\rm d} z$. I chose to replace the ray slopes with the direction cosines relative to the coordinate axes, respectively $s$ and $t$, in order to allow an easier and more elegant formulation of the Snell's law at a surface but at the expense of rendering the translation transformation non-linear.

This work details a general method that can be used to establish all the coefficients needed for any order modelling of optical systems built with surfaces of unspecified shape. The power of the method is exemplified with determination of matrix coefficients for spherical systems in seventh order and for a toroidal surface in third order.

\section{The choice of coordinates}

The optical axis is assumed to lie along the $z$ axis, so the position of any point is determined by the two coordinates $x$ and $y$; when dealing with axis symmetric systems, the two coordinates can be combined in the complex coordinate $X=x + i y$. The ray orientation is defined by cosines of the angles with the $x$ and $y$ axes, respectively $s$ and $t$; again a complex orientation coordinate can be used, $S=s + i t$, in order to simplify the rotational treatment of axis symmetric systems.

Using the same notation of reference \cite{Kondo}, ${\bf x}\&$ is the vector whose components are the coordinates and their monomials up to order $n$:
\begin{equation}
\label{eq:x&}
    {\bf x}\& =
    \left( \begin{array}{c}
        x\\ y\\ s\\ t\\ x^3\\ x^2 y\\ x^2 s\\ x^2 t\\x y^2\\ ...\\ x^j y^k s^l t^m\\ ...
    \end{array} \right)~,
\end{equation}
where total order $j+k+l+m$ is $n$ or less and odd, with all exponents greater than zero. By convention the vector elements are placed in the order of smaller $n$ and larger four digit number $jklm$.

Any transformation of ${\bf x}\&$ into ${\bf x'}\&$ can be represented by a square matrix ${\bf S}$ of dimension $N$ equal to the size of ${\bf x}\&$, following the equation:
\begin{equation}
    \label{eq:transform}
    {\bf x'}\& = {\bf S}{\bf x}\&~.
\end{equation}

The matrix ${\bf S}$ has the following structure:
\begin{equation}
    {\bf S}=
    \left(\begin{array}{cc}
        {\bf P_{4 \times 4}} & {\bf H_{4 \times (N-4)}}\\
        {\bf 0_{(N-4) \times 4}} & {\bf E_{(N-4) \times (N-4)}}
    \end{array} \right)~.
\end{equation}
The four submatrices that form ${\bf S}$ have the following meaning: ${\bf P_{4 \times 4}}$ contains the paraxial constants of the optical system, ${\bf H_{4 \times (N-4)}}$ has $4(N-4)$ high order coefficients from the series expansion of the transformed coordinates, ${\bf 0_{(N-4) \times 4}}$ is composed of zeroes and finally ${\bf E_{(N-4) \times (N-4)}}$ is obtained from the elements of  the first four lines by a procedure called extension whereby the elements of lines 5 to $N$ from ${\bf x'}\&$ are calculated and all terms of orders higher than $n$ are dropped. More specifically, if we wish to determine the coefficients for the line corresponding to the monomial $x^j y^k s^l t^m$, we take the polynomial expressions for $x$, $y$, $s$ and $t$ from lines 1 to 4 and raise them to the powers $j$, $k$, $l$ and $m$, respectively, making their product afterwards; the result is a polynomial of degree $n \times \left(j + k + l + m \right)$ from which only the terms of orders up to $n$ should be considered. The submatrix ${\bf E_{(N-4) \times (N-4)}}$ can itself be subdivided into components of different orders and components with all zero elements.

Although the ray transformation is described by an $N \times N$ matrix, the above considerations show that only the $4(N-4)$ elements from the first four lines need be considered, as the extension procedure is standard for all transformations. In cases where symmetry exists the number of independent coefficients can be greatly reduced \cite{Kondo,Almeida}. It is apparent that the matrix for any given transformation can be considered completely defined when the first four lines' coefficients have been evaluated, i.e. when the transformed coordinates have been expressed in a power series of order $n$.

The coordinate conventions made above are not the same as those made in references \cite{Kondo,Laks} and indeed by the majority of authors; they may appear to be a poorer choice because, as we shall see, they lead to a non-linear translation transformation; on the other hand they will simplify the determination of refraction coefficients. It should be noted, however, that a coordinate change can be seen as a transformation, itself governed by a matrix; it is not difficult to follow the procedure outlined below and then change from cosines into slopes, if needed.

\section{The elementary transformations}
\subsection{The translation matrix}

The first elementary transformation that has to be considered is just the translation of the ray in an homogeneous medium; the orientation coordinates don't change but the position coordinates change according to the equations
\begin{eqnarray}
    x'&=& x + {\frac{s e}{ \sqrt{1 - s^2 -
     t^2}}}~,\nonumber\\
    y'&=& y + {\frac{t e}{ \sqrt{1 - s^2 -
     t^2}}}~,
\end{eqnarray}
where $e$ is the distance travelled, measured along the optical axis.

The series expansion of the equations is rather straightforward; up to the seventh order it is:
\begin{eqnarray}
x'&=&  x + e s + {\frac{e}{ 2}} s^3 + {\frac{e}{ 2}} s t^2 +
    {\frac{3 e}{8} s^5} + {\frac{6 e}{8} s^3 t^2} +
    {\frac{3 e}{8} s t^4}\nonumber\\
    &&+{\frac{5 e}{16} s^7}+ {\frac{15 e}{16} s^5 t^2} +
    {\frac{15 e}{16} s^3 t^4} +
    {\frac{5 e}{16} s t^6}\nonumber\\
y'&=&  y + e t + {\frac{e}{ 2}} s^2 t + {\frac{e}{ 2}} t^3 +
    {\frac{3 e}{8} s^4 t} + {\frac{6 e}{8} s^2 t^3} +
    {\frac{3 e}{8} t^5}\nonumber\\
    &&+{\frac{5 e}{16} s^6 t} +
    {\frac{15 e}{16} s^4 t^3}+
    {\frac{15 e}{16} s^2 t^5} + {\frac{5 e}{16} t^7}~.
\end{eqnarray}

The previous equations give directly the coefficients for lines 1 and 2 of the matrix ${\bf T}$ which describes the translation transformation; lines 3 and 4 are made up of zeroes, except for the diagonal elements which are 1; this translates the two relations $s' = s$ and $t' = t$. The lines 5 to $N$ can be obtained by the extension procedure described before.

If slopes were used instead of direction cosines the translation matrix would have a much simpler form, corresponding to a linear transformation \cite{Kondo}.

\subsection{Refraction formulae}

The influence of the surface power on the ray coordinates must now be considered. It is useful to analyse separately the changes on the ray orientation and the modification of the position coordinates. The ray orientation changes due to the application of Snell's law when the ray encounters the surface of separation between the two media; let this surface be defined by the general formula:
\begin{equation}
\label{eq:surface}
{\rm f}(x,y,z) = 0
\end{equation}

Let a ray impinge on the surface on a point of coordinates $x$ and $y$ with direction cosines $s$ and $t$. The normal to the surface at the incidence point is the vector:
\begin{equation}
    \mathbf{n} = {\bf grad}\,{\rm f}~;
\end{equation}
the direction of the incident ray is the unit vector:
\begin{equation}
\label{eq:40}
    \mathbf{v} =
    \left(\begin{array}{c}
        s \\
        t \\
        \sqrt{1 - s^2 - t^2}
    \end{array}\right)~;
\end{equation}
and the refracted ray has a direction which is represented by the vector:
\begin{equation}
\label{eq:41}
    \mathbf{v'} =
    \left(\begin{array}{c}
        s' \\
        t' \\
        \sqrt{1 - {s'}^2 - {t'}^2}
    \end{array}\right)~,
\end{equation}
where $s'$ and $t'$ are the direction cosines after refraction.

One can apply Snell's law by equating the cross products of the ray directions with the normal on both sides of the surface multiplied by the respective refractive indices:
\begin{equation}
\label{eq:snell}
\nu \,{\bf v} \otimes {\bf n} = {\bf v'} \otimes {\bf n}~,
\end{equation}
where $\nu$ represents the refractive index ratio of the two media.

The algebraic solutions of the equation above can be found for many surfaces by means of suitable symbolic processing software like Mathematica; in certain cases the solution is easier to find after a suitable change of coordinates along the $z$ axis, which will not affect the final result; for instance, for a spherical surface it is convenient to shift the coordinate origin to the center. In order to find the coefficients for the matrix ${\bf R_1}$, representing the change in the ray orientation, the explicit expressions for $s'$ and $t'$ must be expanded in series; this can again be programmed into Mathematica. The examples given later show this procedure for two different surfaces.

\subsection{Surface offset}

The use of just two position coordinates implies that these are referenced to a plane normal to the axis. When a ray is refracted at a surface, the position coordinates that apply are those of the incidence point; nevertheless the translation to the surface is referenced to the vertex plane and thus the position coordinates that come out of the translation transformation, are those of the intersection with that plane; the difference between the two positions is named surface offset.

\begin{figure}[htb]
    \centerline{\psfig{file=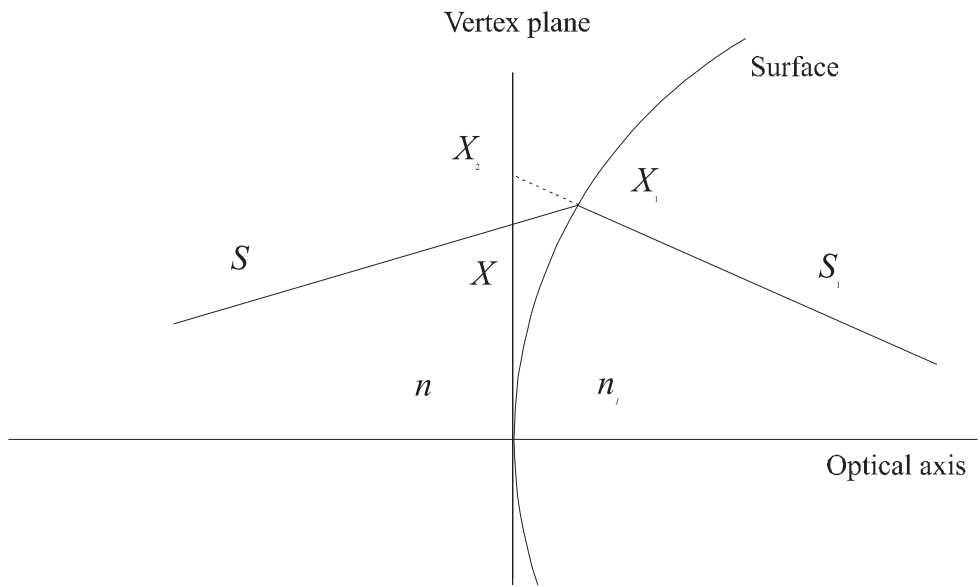, scale=1.1}}
\caption{\label{offset} The ray intersects the surface at a point
$X_1$ which is different both from the point of intersection of the
incident ray with the plane of the vertex, $X$, and the point of
intersection of the refracted ray with the same plane, $X_2$. The
surface is responsible for three successive transformations: 1 - an
offset from $X$ to $X_1$, 2 - the refraction and 3 - the offset from
$X_1$ to $X_2$.}

\end{figure}

The figure \ref{offset} illustrates the problem that must be solved: The reference plane for the position coordinates of the ray in a refraction process is the plane of the surface vertex; the coordinates of the point of intersection of the ray with this plane are not the same before and after the refraction. The surface matrix must account not only for the orientation changes but also for the position offset introduced by the refraction process.

If the origin of the rectangular coordinates is located at the vertex and $(x,y,0)$ are the coordinates of the ray when it crosses the plane of the vertex, the current coordinates of a point on the incident ray are given by the two equations:
\begin{eqnarray}
\label{eq:offset1}
    x'&=& x + {\frac{s z'}{ \sqrt{1 - s^2 -
     t^2}}}~;\nonumber\\
    y'&=& y + {\frac{t z'}{ \sqrt{1 - s^2 -
     t^2}}}~.
\end{eqnarray}

As the intersection point has to verify both (\ref{eq:offset1}) and (\ref{eq:surface}) it is possible to solve a system of three simultaneous equations to determine its coordinates. In fact there may be several intersection points, as the ray may intercept a complex surface in several points; it is not difficult, however, to select the solution of interest by careful examination of the surface region that has been hit; we are usually interested in the solution that has the smallest $\left| z'\right |$ value.

As before a series expansion of the exact expressions of $x'$ and $y'$ must be performed taking only the terms up to the $n$th order; this will yield the coefficients for the matrix ${\bf T_1}$ which describes the transformation from point $X$ to point $X_1$ and will be called the forward offset.

The offset from point $X_1$ to point $X_2$, designated backward offset, is the reverse transformation of the forward offset, for a ray whose direction coordinates have been modified by refraction. We solve (\ref{eq:offset1}) and (\ref{eq:surface}) in terms of $x$ and $y$ and apply the previous procedure to determine the coefficients of the transformation matrix, ${\bf T_2}$.

The matrix describing the transformations imposed by the surface is the product ${\bf T_2}{\bf R_1}{\bf T_1}$, as the ray has to suffer the three successive transformations: forward offset, refraction and backward offset. The surface matrix is called ${\bf R}$.

\section{Stop effects}

The entrance pupil of the optical system plays an important role in the overall aberrations; the center of the pupil defines the chief ray, which determines the paraxial image point. The ray fan usually extends equally in all directions around the chief ray and the points of intersection of the various rays in the fan with the image plane determine the ray aberrations. When matrices are used to model the system the image appears described in terms of the position and orientation coordinates of the rays when these are subjected to the first transformation, be it a refraction or a translation; in terms of ray aberrations this corresponds to a situation where the object is located at infinity and the entrance pupil is placed where the first transformation takes place. In fact, the orientation coordinates play the role of object coordinates and the position coordinates are the actual pupil coordinates, if this coincides with the first transformation; the chief ray can be found easily just by zeroing the position coordinate.

Ray aberrations can be evaluated correctly if the image is described in terms of both object and stop coordinates or some appropriate substitutes; an object point is then set by the object coordinates, the paraxial image point is found when the stop coordinate is zero and the aberrations are described by the differences to this point. This can be done adequately when the stop is located before the first surface and so coincides with the entrance pupil. If the object is at infinity the problem is very easy to solve and application examples have already been described \cite{Almeida}. If the entrance pupil is located at the position $z_s$ relative to the first surface, this means that the rays must perform a translation of $-z_s$ prior to hitting it; this can be accommodated via the right product by a translation matrix ${\bf T_{-z_s}}$, where the index indicates the amount of translation.

\begin{figure} [htb]
    \centerline{\psfig{file=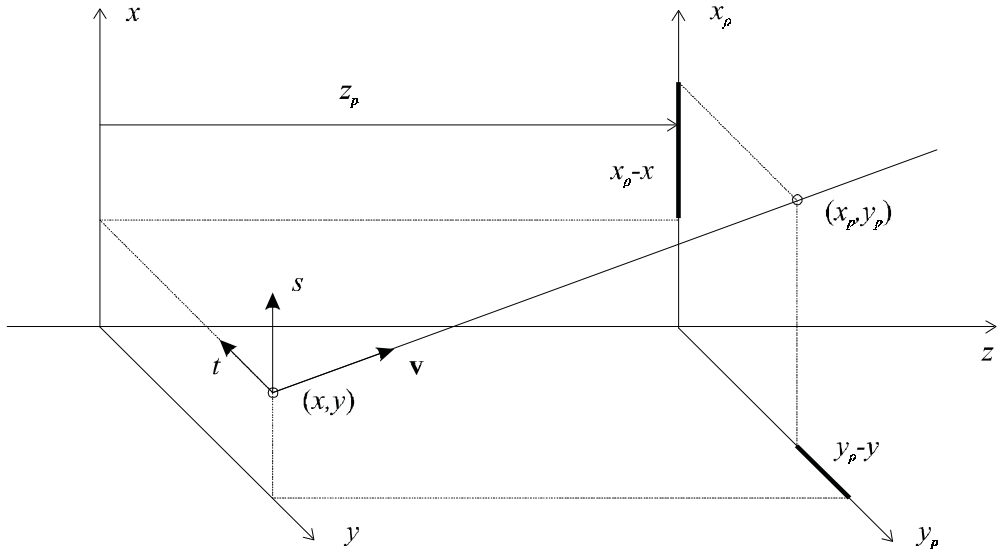, scale=0.8}}
\caption{\label{pupil} The coordinates of the point where the ray
crosses the entrance pupil, together with the object point
coordinates are a convenient way of defining the ray orientation in
order to incorporate stop effects.}

\end{figure}

The case of near objects is more difficult to deal with because we
are faced with a dilemma: If a translation from the pupil position
to the first surface is applied the position coordinates of the
incident rays become pupil coordinates but the rays' origins on the
object are lost. Conversely we can apply a translation from the
object position, ending up with the reverse problem of knowing the
rays' origins and ignoring their point of passage through the pupil.
We need the position coordinates both at the object and pupil
location; this suggests that the ray's orientation should be
specified not by its direction cosines but by the coordinates of the
point of passage through the pupil. In figure \ref{pupil} a ray
leaves an object point of coordinates $(x,y)$ and crosses the
entrance pupil plane at a point of coordinates $(x_p,y_p)$; then the
direction cosines, $s$ and $t$ can be calculated by:
\begin{eqnarray}
\label{eq:pupil}
s &=& \frac{x_p - x}{\sqrt{(x_p - x)^2 + z_p^2}}~, \nonumber\\
t &=& \frac{y_p - y}{\sqrt{(y_p - y)^2 + z_p^2}}~,
\end{eqnarray}
where $z_p$ is the position of the pupil relative to the object.

The matrix theory developed before was based on direction cosines rather than pupil coordinates; in consequence a coordinate change is needed before the translation from the object to the first surface is applied to the rays. Rather than use equations (\ref{eq:pupil}) it is more convenient to look at them as a  coordinate change governed by a matrix ${\bf T_p}$; for the determination of its coefficients we apply the standard procedure of series expansion. The following is the result for 7th order:
\begin{eqnarray}
s &=& -\frac{1}{z_p} x +\frac{1}{z_p} x_p + \frac{1}{2 z_p^3} x^3
    -\frac{3}{2 z_p^3} x^2 x_p +\frac{3}{2 z_p^3}x x_p^2
    -\frac{1}{2 z_p^3}x_p^3 -\frac{3}{8 z_p^5} x^5
    +\frac{15}{8 z_p^5}x^4 x_p \nonumber\\
    &&-\frac{15}{4 z_p^5}x^3 x_p^2 +\frac{15}{4 z_p^5}x^2 x_p^3
    -\frac{15}{8 z_p^5}x x_p^4
    +\frac{3}{8 z_p^5} x_p^5 +\frac{5}{16 z_p^7} x^7
    -\frac{35}{16 z_p^7}x^6 x_p +\frac{105}{16 z_p^7}x^5 x_p^2 \nonumber\\
    &&-\frac{175}{16 z_p^7}x^4 x_p^3 +\frac{175}{16 z_p^7}x^3 x_p^4
    -\frac{105}{16 z_p^7}x^2 x_p^5 +\frac{35}{16 z_p^7}x x_p^6
    -\frac{5}{16 z_p^7}x_p^7~;
\end{eqnarray}
a similar expression exists for $t$. The equation above allows the determination of the coefficients for matrix ${\bf T_p}$ which converts the coordinates $\left(x, y, x_p, y_p \right)$ into $\left(x, y, s, t \right)$ before the rays enter the system at the object position; as a symmetrical conversion is not performed when the rays leave the system, there is no place for the use of matrix ${\bf T_p^{-1}}$, as would be the case in a similarity transformation.

A case study would involve the following successive steps:
\begin{itemize}
\item Determine the matrices for all the translation and refraction transformations suffered by the ray from the object point to the image plane.
\item Determine the matrix ${\bf T_p}$, relative to the entrance pupil position.
\item Multiply all the matrices in reverse order, starting with the translation from the last surface to the image plane and ending with matrix ${\bf T_p}$; the result is matrix ${\bf S}$.
\item Determine a vector base ${\bf x}\&$ with the set of coordinates $\left(x,y,x_p,y_p \right)$.
\item Make the product ${\bf x'}\& = {\bf S} {\bf x}\&$.
\end{itemize}

Once ${\bf x'}\&$ has been calculated, its first four elements are polynomials of $n$th degree on the independent variables $\left(x,y,x_p,y_p \right)$, which model the position and direction cosines of the rays on the image plane.

Most optical systems have stops located after the first surface, and so the method described above is not adequate. One can always find the gaussian entrance pupil of any system and so revert to the previous situation; this will work even if the entrance pupil is found to lie beyond the first surface, in which case it will give rise to a negative translation. The problem is that the gaussian entrance pupil is a first order transformation of the stop, whereas the real entrance pupil is an aberrated image of the stop given by the portion of the optical system that lies before it. Ideally one should try to describe the image in terms of the object coordinates and the point of passage through the stop, which can be rather difficult. Alternatively one can find a reference ray, defined as the ray that passes through the center of the stop and is not necessarily the same as the chief ray, and work with position and orientation coordinates on the image plane, relative to the reference ray. The plot of the relative ray position versus its relative direction cosines can be used to describe ray aberrations \cite{Goodman}.

\section{Coefficients for a spherical surface}

A sphere of radius $r$ with its center at $z=r$ is defined by the equation:
\begin{equation}
\label{eq:sphere}
x^2 + y^2 + z^2 - 2 z r = 0~;
\end{equation}
this equation will replace (\ref{eq:surface}) in the algorithm for the determination of matrix coefficients. In a previous work \cite{Almeida} we presented the results for third order; these are now extended to seventh order.

This is a situation where symmetry must be considered in order to reduce the size of the matrices involved; we use the complex variables $X$ and $S$ previously defined, leading to a complex vector base ${\bf X}\&$ whose elements have the general form $X^j X^{*k} S^l S^{*m}$ with $X^*$ and $S^*$ representing the complex conjugates of $X$ and $S$. Kondo et al. \cite{Kondo} have shown that the powers must obey the condition $j-k+l-m=1$; for seventh order the allowable combinations are:
\begin{eqnarray} &&1000,0010,2100,2001,1110,1011,0120,
    0021,3200,3101,3002,2210,2111,2012,1220,1121,\nonumber\\
    &&1022,0230,0131,0032,4300,4201,4102,4003,3310,
    3211,3112,3013,2320,2221,2122,2023,\nonumber\\
    &&1330,1231,1132,1033,0340,0241,0142,0043 ~.\nonumber
\end{eqnarray}

Equation (\ref{eq:snell}) can now be solved and the resulting expressions for $s'$ and $t'$ can be expanded in series up to the seventh order; the results can be combined into one complex variable $S'$. The second column of table \ref{t:spherical} lists the coefficients for this expansion. Similarly the forward offset coefficients can  be found solving equations (\ref{eq:offset1}) and combining $x'$ and $y'$ expansions into one single complex variable $X_1$; the resulting coefficients are listed in column 3 of table \ref{t:spherical}. The backward offset coefficients are obtained in a similar way, although now it is the $x$ and $y$ expansions that must be combined into a single complex variable $X_2$.

\section{Coefficients for an asymmetric surface}

Toroidal surfaces are used to correct eye astigmatism; they can be
fabricated easily without resorting to sophisticated machinery. In
this example we deem to show that the method is applicable to
general surfaces but it is not our purpose to list high order
coefficients for this particular shape; therefore we will restrict
the study to third order. Also, as there is no rotational symmetry
we will not use complex coordinates; there is symmetry relative to
two perpendicular planes and if we were to use complex coordinates
the terms should obey the condition  $j-k+l-m= \pm 1$ as explained
in reference  \cite{Laks}; we feel that it is clearer not to invoke
symmetry at all.

The surface is generated by a circle of radius $r_1$ normal to the $x z$ plane whose center is displaced along a circle of radius $r_2$ laying on this plane. This surface has curvature radii of $r_1$ and $r_1 + r_2$ respectively on the $x z$ and $y z$ planes. The surface equation can be written as:
\begin{equation}
\label{eq:toroid}
\left(x^2+y^2+z^2-r_1^2-r_2^2\right)^2-4 r_2^2\left(r_1^2-x^2\right) =0~.
\end{equation}

The solution of equation (\ref{eq:snell}) originates the expressions for $s'$ and $t'$; these are then expanded in series up to the third order and the resulting coefficients are listed in columns 2 and 3 of table \ref{t:toroid}. The same procedure was applied to the expressions for $x'$ and $y'$ resulting from equation (\ref{eq:offset1}); columns 4 and 5 of the same table list these coefficients. The backward offset coefficients need not be listed, as in third order they can be obtained by sign reversal from the forward offset coefficients.

\section{Discussion and conclusion}

Matrices can be used to model optical systems built with surfaces of unspecified shape to any desired degree of approximation. The method that was described differs from those found in literature by the choice of direction cosines rather than ray slopes to specify ray orientations. This results in a very elegant formulation of Snell's law which can be easily programmed into symbolic computation software. In spite of some added complication in translation matrices the algorithm is limited only by computing power in it's ability to determine the matrix coefficients, irrespective of surface complexity and order of approximation. For axis-symmetric systems the matrices have a dimension $40 \times 40$ for seventh order and can usually be dealt with by ordinary PCs; the influence of stops can also be incorporated in the matrix description as was demonstrated. The unconventional coordinate system that was used can, if needed, be converted to the more usual ray slope coordinate system through the product with conversion matrices of the same dimension.

Two examples were shown for illustration of the method's capabilities; the first one supplies a list of seventh order coefficients for systems made up with spherical surfaces and allows the reader an immediate use. The second example applies the method to a fourth degree toroidal surface, showing that complex shapes can be accommodated.

\begin{table}
\caption{7th order coefficients for spherical surfaces}
\begin{tabular}{llll}
\hline
Monomial & Refraction $(S')$ & Forward offset $(X_1)$ & Backward offset $(X_2)$\\
\hline
$X$ & $(\nu-1)/r$& $1$ & $1$ \\
$S$ & $\nu$& $0$ & $0$ \\
$X^2 X^*$ & $\nu(\nu-1)/2 r^3$& $0$ & $0$ \\
$X^2 S^*$ & $\nu(\nu-1)/2 r^2$& $0$ & $0$ \\
$X X^* S$ & $\nu(\nu-1)/2 r^2$& $1/2 r$ & $-1/2 r$ \\
$X S S^*$ & $\nu(\nu-1)/2 r$& $0$ & $0$\\
$X^* S^2$ & $0$& $0$ & $0$\\
$S^2 S^*$ & $0$& $0$ & $0$\\
$X^3 X^{*2}$ & $\nu(\nu^3 - 1)/8 r^5$& $0$ & $0$\\
$X^3 X^* S^*$ & $\nu^2(\nu^2 - 1)/4 r^4$& $0$ & $0$\\
$X^3 S^{*2}$ & $\nu^2(\nu^2 -1)/8 r^3$& $0$ & $0$\\
$X^2 X^{*2} S$ & $\nu^2(\nu^2 -1)/4 r^4$& $1/8 r^3$ & $-1/8 r^3$\\
$X^2 X^* S S^*$ & $\nu(2 \nu^3 - 3\nu +1)/4 r^3$& $0$ & $1/4 r^2$\\
$X^2 S S^{*2}$ & $\nu^2(\nu^2 -1)/4 r^2$& $0$ &$0$ \\
$X X^{*2} S^2$ & $\nu^2(\nu^2 -1)/8 r^3$& $0$ & $1/4 r^2$ \\
$X X^* S^2 S^*$ & $\nu^2(\nu^2 -1)/4 r^2$& $1/4 r$ & $-1/4 r$ \\
$X S^2 S^{*2}$ & $\nu(\nu^3 -1)/8 r$& $0$ &$0$ \\
$X^{*2} S^3$ & $0$& $0$ &$0$ \\
$X^* S^3 S^*$ & $0$& $0$ &$0$ \\
$S^3 S^{*2}$ & $0$& $0$ &$0$ \\
$X^4 X^{*3}$ & $\nu(\nu^5 -1)/16 r^7$& $0$ &$0$ \\
$X^4 X^{*2} S^*$ & $\nu^2(3 \nu^4 -2\nu^2 -1)/16 r^6$& $0$ &$0$ \\
$X^4 X^* S^{*2}$ & $3\nu^4(\nu^2 -1)/16 r^5$& $0$ &$0$ \\
$X^4 S^{*3}$ & $\nu^4(\nu^2-1)/16 r^4$& $0$ &$0$ \\
$X^3 X^{*3} S$ & $\nu^2(3\nu^4 -2\nu^2 -1)/16 r^6$& $1/16 r^5$ &$-1/16 r^5$ \\
$X^3 X^{*2} S S^*$ & $\nu(9 \nu^5 - 10 \nu^3 +1)/16 r^5$&$0$ & $3/16 r^4$ \\
$X^3 X^* S S^{*2}$ & $\nu^2(9\nu^4 - 11\nu^2 +2)/16 r^4$&$0$ & $-1/8 r^3$ \\
$X^3 S S^{*3}$ & $3\nu^4(\nu^2-1)/16 r^3$& $0$ &$0$ \\
$X^2 X^{*3} S^2$ & $3\nu^4(\nu^2 -1)/16 r^5$&$0$ & $3/16 r^4$ \\
$X^2 X^{*2} S^2 S^*$ & $\nu^2(9\nu^4 -11\nu^2 +2)/16 r^4$& $1/16 r^3$ &$-7/16 r^3$ \\
$X^2 X^* S^2 S^{*2}$ & $\nu(9\nu^5 -10\nu^3 +1)/16 r^3$&$0$ & $1/4 r^2$ \\
$X^2 S^2 S^{*3}$ & $\nu^2 (3 \nu^4 - 2 \nu^2 -1)/16 r^2$& $0$ &$0$ \\
$X X^{*3} S^3$ & $\nu^4(\nu^2-1)/16 r^4$&$0$ & $-1/8 r^3$ \\
$X X^{*2} S^3 S^*$ & $3\nu^4(\nu^2-1)/16 r^3$&$0$ & $1/4 r^2$ \\
$X X^* S^3 S^{*2}$ & $\nu^2(3\nu^4 -2\nu^2 -1)/16 r^2$& $3/16 r$ &$-3/16 r$ \\
$X S^3 S^{*3}$ & $\nu(\nu^5 -1)/16 r$& $0$ &$0$ \\
$X^{*3} S^4$ & $0$& $0$ &$0$ \\
$X^{*2} S^4 S^*$ & $0$& $0$ &$0$ \\
$X^* S^4 S^{*2}$ & $0$& $0$ &$0$ \\
$S^4 S^{*3}$ & $0$& $0$ &$0$
\end{tabular}
\label{t:spherical}
\end{table}

\begin{table}
\caption{3rd order coefficients for a torous}
\begin{tabular}{lllll}
\hline
Monomial & Refraction $(s')$ & Refraction $(t')$ & Forward offset $(x')$ &  Forward offset $(y')$\\
\hline
$x$ & $\frac{\nu-1}{r_1}$& $0$ & $1$ & $0$\\
$y$ & $0$& $\frac{\nu-1}{r_1 + r_2}$ & $0$ & $1$\\
$s$ & $0$& $0$ & $0$ & $0$\\
$t$ & $0$& $0$ & $0$ & $0$\\
$x^3$ & $\frac{\nu(\nu-1)}{2 r_1^3}$& $0$ & $0$ & $0$\\
$x^2 y$ & $0$& $\frac{\nu(\nu-1)}{2 r_1^2 (r_1 + r_2)}$ & $0$ & $0$\\
$x^2 s$ & $\frac{\nu(\nu-1)}{r_1^2}$& $0$ & $\frac{1}{2 r_1}$ & $0$\\
$x^2 t$ & $0$& $0$ & $\frac{1}{2 r_1}$ & $0$\\
$x y^2$ & $\frac{\nu(\nu-1)}{2 r_1 (r_1 + r_2)^2}$& $0$ & $0$ & $0$\\
$x y s$ & $0$& $\frac{\nu(\nu-1)}{r_1(r_1 + r_2)}$ & $0$ & $0$\\
$x y t$ & $\frac{\nu(\nu-1)}{r_1(r_1 + r_2)}$& $0$ & $0$ & $0$\\
$x s^2$ & $\frac{\nu(\nu-1)}{2 r_1}$& $0$ & $0$ & $0$\\
$x s t$ & $0$& $0$ & $0$ & $0$\\
$x t^2$ & $\frac{\nu(\nu-1)}{2 r_1}$& $0$ & $0$ & $0$\\
$y^3$ & $0$& $\frac{\nu(\nu-1)}{2(r_1 + r_2)^3}$ & $0$ & $0$\\
$y^2 s$ & $0$& $0$ & $0$ & $\frac{1}{2 (r_1+r_2)}$\\
$y^2 t$ & $0$& $\frac{\nu(\nu-1)}{(r_1 + r_2)^2}$ & $0$ & $\frac{1}{2(r_1+r_2)}$\\
$y s^2$ & $0$& $\frac{\nu(\nu-1)}{2 (r_1 + r_2)}$ & $0$ & $0$\\
$y s t$ & $0$& $0$ & $0$ & $0$\\
$y t^2$ & $0$& $\frac{\nu(\nu-1)}{2 (r_1 + r_2)}$ & $0$ & $0$\\
$s^3$ & $0$& $0$ & $0$ & $0$ \\
$s^2 t$ & $0$& $0$ & $0$ & $0$\\
$s t^2$ & $0$& $0$ & $0$ & $0$\\
$t^3$& $0$& $0$ & $0$ & $0$
\end{tabular}
\label{t:toroid}
\end{table}


\begin{thebibliography}{22}
% Please use the \bibitem command to create references.
\bibitem{Gerrard}A. Gerrard, J.M. Burch, {\it Introduction to Matrix
Methods in Optics} (Dover Publications, New York, 1994).

\bibitem{Born}M. Born and E. Wolf, {\it Principles of Optics}
(Pergamon Press, N.Y., 1980).

\bibitem{Slyusarev}G.G. Slyusarev, {\it Aberration and Optical Design
Theory} (Adam Hilger Ltd., Bristol, 1984).

\bibitem{Kondo}M. Kondo and Y. Takeuchi, "Matrix method for nonlinear
transformation and its application to an optical lens system", J.
Opt. Soc. Am. A {\bf 13}, 71-89, (1996).

\bibitem{Laks}V. Lakshminarayanam and S. Varadharajan, "Expressions
for aberration coefficients using nonlinear transforms", Optom. and
Vision Sci. {\bf 74}, 676-686, (1997).

\bibitem{Almeida}J.B. Almeida, "The use of matrices for third order
modeling of optical systems", in {\it International Optical Design
Conference}, K.P. Thompson and L.R. Gardner, eds., Proc. SPIE {\bf
3482}, 917-925, (1998).

\bibitem{Goodman}D.S. Goodman, in {\it Handbook of Optics}, M. Bass,
ed. in chief, (Mc. Graw-Hill, N.Y., 1995), Vol. 1, p. 93.

\end{thebibliography}
\end{document}